# Turnover of the melting line of highly compressed molecular hydrogen

Eugene Yakub [*)]

*Cybernetics Dept., Odessa National Economic University, 65082, Preobrazhenskaya 8, Odessa, Ukraine*

Conventional molecular dynamics simulation has been used to determine melting temperature of highly compressed classical molecular hydrogen in a wide range of pressures and temperatures using non-empirical atom-atom potentials approximation. It is shown that the raise of melting temperature with increasing pressure becomes more and more smooth and at megabar pressures its slope goes negative. We discuss the reasons of this turnover and analyze connection of this effect with the parameters of chemical bonding in $H_2$.

(Dated March 15, 2013)

## I. INTRODUCTION

Turnover in the melting line of molecular hydrogen has been intensively discussed in the literature during last years[1-4]. Experimentalists worked hard to verify the existence of a maximum of melting temperature as a function of pressure in compressed hydrogen predicted in *ab initio* simulations[1]. Despite the serious difficulties in measurements at megabar pressures and elevated temperatures, several research groups have confirmed the existence of the turnover in the melting line[2-4].

There are different points of view on what the physics is behind this effect, in particular, specific changes in the intermolecular interaction have been assumed as its reason[1]. However, discovery of similar maxima of melting temperature in other simple molecular systems, particularly in carbon, water and nitrogen[5], give some grounds to think that the origin of the turnover effect may be more general.

---

[*] Electronic mail: yakub@oneu.edu.ua

The aim of the present work is study of melting in highly compressed classical molecular hydrogen $T_m(P)$ within the framework of a non-empirical atom - atom potentials (AAP) approximation[7]. We applied the conventional single-box molecular dynamics (MD) and compared the results of our simulations with *ab initio* predictions[1] and existing experimental data[2-4].

In the next section we describe briefly basics and practical implementation of AAP model for molecular hydrogen. In Section III we provide details of MD simulation performed. The simulation results their comparison and discussion are presented in Section IV, and general conclusions are formulated in the last section.

## II. ATOM- ATOM POTENTIALS

Interaction potentials of hydrogen molecules were subject of many studies (see *e.g.* recent paper of Freiman *ea.*[6] and references therein). Most of them have been based on semi-empirical isotropic potentials, which consider the anisotropy of intermolecular interaction as a small correction and ignore interplay between inter- and intramolecular coordinates. This approach is quite reasonable at low temperatures and densities where molecular excitations play a minor role. A relatively simple AAP model (see *e.g.* Ref. 7) stands somewhat apart. It has been proposed a long ago and was applied in studies of a wide spectrum of problems, where rotational and vibrational excitations of diatomic molecules are crucial: from evaluating of the third virial coefficient of dissociating hydrogen up to predicting structure, thermodynamic and transport properties, polymorph transitions in solid $H_2$ and Hugoniot curves of fluid hydrogen[7-11].

First theoretical justification of AAP approximation, based on expansion of London's potential surfaces of four hydrogen atoms given within valence bond theory[11], was later reformulated in much more general form[7] for arbitrary systems of diatomic molecules in their singlet ground states within the framework of molecular orbital method (Böhm-Ahlrichs theorem[12]).

According to AAP model, potential energy of a system of *N* homonuclear diatomics can be represented as a function of interatomic distances of all their 2*N* constituent atoms:

$$U_{2N} = \sum_{(ij)} u(r_{ij}) + \sum_{i<j} \Phi(r_{ij}). \qquad (1)$$

Here $u(r_{ij})$ is the potential energy of a diatomic molecule as a function of internuclear distance (*i.e.* intramolecular interaction potential of *i*-th and *j*-th atoms forming this molecule), and $\Phi(r_{ij})$ is atom-atom potential describing interactions between atoms bound in different molecules. Correspondingly, the first sum in Eq.(1) extends over all *N* molecules, and the second is taken over all $4N(N-1)$ pairs of atoms belonging to different molecules.

According to the above mentioned Böhm-Ahlrichs theorem[12], the non-bonding atom-atom potential $\Phi(r_{ij})$ can be expressed as a weighted average of interaction energies of two isolated atoms over all their electronic terms. In the particular case of molecular hydrogen the intramolecular potential $u(r)$ corresponds to the ground (singlet) state of $H_2$ molecule and intermolecular atom-atom potential $\Phi(r)$ is the sum of ¼ of singlet $u(^1\Sigma|r)$ and ¾ of triplet $u(^3\Sigma|r)$ potential curves of $H_2$ molecule:

$$\Phi(r) = \tfrac{1}{4} u(^1\Sigma|r) + \tfrac{3}{4} u(^3\Sigma|r). \qquad (2)$$

Both singlet and triplet potential curves are well known since classical *ab initio* variational calculations of Kolos and Wolniewitz[13]. In this work we applied AAP model in the same form as in our previous studies[7-11] aiming to investigate the location and shape of the melting curve $T_m(P)$ in classical highly compressed molecular hydrogen. Potential energy of the $H_2$ singlet ground state $u(^1\Sigma|r)$ was approximated by the modified Hulburt-Hirschfelder potential[11]:

$$u(^1\Sigma|r) = D_e \left[ e^{-2x} - 2e^{-x} - ax^3(1-bx)e^{-cx} \right], \qquad (3)$$

where $x = b(r/r_e - 1)$, $r_e$ = 0.74126 Å, $D_e$ = 4.747 eV, $b = 1.4403$, a=0.1156, b=1.0215, c=1.72.

Eq. (3) provides an excellent approximation of the $^1\Sigma_g^+$-curve within a wide range of distances (0.3…5Å). Non-bonding atom-atom potential $\Phi(r)$ was represented by the following form proposed by Saumon and Chabrier[14]:

$$\Phi(r) = \varepsilon\left[\gamma e^{-2s_1(r-r^*)} - (1+\gamma)e^{-s_2(r-r^*)}\right]. \quad (4)$$

Parameters $r^*$ = 3.2909 Å, $\varepsilon$ = 1.74×10$^{-3}$ eV, $\gamma$ = 0.4615, $s_1$ = 1.6367 Å$^{-1}$, and $s_2$ = 1.2041 Å$^{-1}$, were obtained in Ref. 14 on the basis of *ab initio* calculations of Kolos and Wolniewitz[13].

Eq.(4) gives an accurate description of non-bonding potential Eq.(2) over a wide interval of distances (0.5… 6.5 Å), including the region of strong repulsion at short distances and the region of weak dispersional attraction at large distances.

### III. MD SIMULATION

To locate the melting line of classical molecular $H_2$ solid in the *T-P* diagram within AAP approximation, we performed a series of conventional single-box Nose-Hoover *N-V-T* MD simulations at high densities and temperatures above 300 K (see Table 1 for technical details).

To keep an eye on possible effect of periodical boundary conditions (PBC) on parameters of phase transition inside a finite MD cubic box all simulations were carried out using two different sizes of MD box, containing 512 atoms (256 molecules) or 1782 atoms (864 molecules).

Every simulation starts from the same initial configuration where molecular centers form a close packed *fcc* crystalline structure and all intramolecular distances correspond to the equilibrium bond length $r_e$ = 0.741 Å. Several types of initial orientations of molecules in such lattice were tested but no effect on the final simulation results was found, because at high temperatures molecular rotations and vibrations are so intense that details of the initial static configuration are quickly forgotten.

Instant values of pressure, energy, mean chemical bond length, and temperature fluctuations have been periodically (every 0.25 *ps* of simulation time) recorded in the listing file for further analysis. Establishing and maintaining of thermodynamic equilibrium was monitored continuously by comparing the actual and Maxwellian velocity distributions on the screen. We evaluated also the impact of the Nose-Hoover thermostat coupling parameter choice to avoid the effect of possible large fluctuations of temperature in small MD boxes and nothing was found.

When estimating melting temperature at a given density, two functions most sensitive to the melting transition were monitored continuously:

1) Center-to-center molecular radial distribution function;
2) Mean squared displacement (MSD) of molecular centers as a function of time.

Additional test of molecular structure was provided by periodical recording of instant positions of all atoms inside the box for later examination using visualization tools provided by JMol java viewer[15].

A series of MD simulations was carried out at several fixed densities ranging from 0.3 to 0.95 g cm$^{-3}$. The temperature at given density was varied step-by-step from one simulation to another by a fixed increment (25 K). Eventually we were able to detect the maximum temperature corresponding to the crystalline state remaining thermodynamically stable during all the simulation period.

It is worth mention here that strictly speaking it is impossible to determine the exact location of the melting line using the single-box MD technique only. In our simulation we were actually able to detect the limit of the lattice stability, which can be regarded as an upper bound of the melting temperature. This limit may exceed the real thermodynamic melting temperature $T_m$ because the crystal can exist in a metastable state at $T > T_m$. Nevertheless the achievable overheating of van-der-waals' crystals is rather small[16] and hardly will exceed the adopted temperature increment (25 K), which finally determines the accuracy of our estimation of $T_m$.

## IV. RESULTS AND DISCUSSION

Results of our MD simulations are presented in Figs. 1 - 3. Fig.1 illustrates differences in the structure and diffusion between solid and liquid hydrogen found at the melting point.

In Fig.2 the predicted melting temperatures are compared with experimental data of Deemyad and Silvera[2], Gregorianz ea.[17], and Eremets and Troyan[3] as well as with *ab initio* simulation data of Bonev *ea.*[1]. In our opinion, the most interesting result here is a pronounced turnover of the melting line predicted in the pressure range 150… 250 GPa.

The secondary effect that can be clearly seen in Fig.2, is the specific shape of the melting line near the maximum of melting temperature. Just after this maximum at 150 GPa, there appears an evident flattening (or even notch) located in that pressure range where the line of the I-III polymorph transition shown in Fig.2 by dash-dotted line according to estimation of Silvera and Deemyad[21], crosses the predicted melting line. Taking into account the fact that AAP approximation actually reproduces the orientational ordering of hydrogen molecules and this I-III transition line in solid phase[10], one may speculate that the right part of the melting line at P>200 GPa corresponds to the melting of a more orientationally ordered phase of solid hydrogen.

No significant effect of PBC was found in our simulations, except of the highest densities studied. At maximal pressures about 350 GPa and in tests on a very long simulation times (>300 *ps*) we observed some kind of recovery of the solid phase after melting in the smaller (216 molecules) MD box. MSD plot here was clearly divided here into three linear sections. The first, almost horizontal one, corresponds to the initial solid structure with a very low diffusion coefficient. After a rather long period of simulation (about 150 *ps*) the slope of the plot increases suddenly about forty times (alike in the inset on Fig.1) indicating melting of the initial crystalline structure. Liquid phase remains stable for the next 170 *ps*, then the diffusion coefficient drops abruptly more than twice and another solid structure, highly defective and diffusive is formed in the box. No such effect was observed in bigger boxes. We attribute it to the artificial periodicity imposed by PBC.

In order to clarify the main question: what the physics is behind the overall turnover of the melting line, we performed a few additional MD experiments varying potential parameters responsible for chemical bonding in AAP model.

The results are presented in Fig.3. The first experiment was intended to shed light on the role of non-spherical part of intermolecular interaction. In the first series of such simulations the equilibrium bond length – parameter $r_e$ in Eq.(3) was shortened by half. The corresponding location of the melting line is shown in Fig.3 by curve 1. Much more steep raise of melting temperature in this case indicate the importance of non-spherical part of short-range molecular interaction in explanation of the melting line turnover effect.

Second and third experiments have been performed to asses the role of the chemical bond rigidity. We estimated the location of the melting line again using the same AAP approximation with changed parameter $b$ in Eq. (3) responsible for stiffness of the chemical bond (vibronic frequency). In the second experiment it was reduced by half (curve 2) and in the third one it was doubled (curve 3).

As one can see, the stiffness of the chemical bond at a fixed its equilibrium length also plays an important role in formation of the melting curves shape. The stiffer is chemical bonding (and, correspondingly, the higher is vibronic frequency), the more pronounced is the effect of turnover.

Although the finding of the volume change on melting is beyond the scope of this work, some qualitative conclusions about it can be made indirectly by analyzing of obtained MD data. It was found that in the whole range of pressures studied, the melting process at a fixed density in MD box was always accompanied by certain increase in internal energy. On the contrary, changes in pressure observed at melting are different at low and high densities. Pressure in MD box slightly decreases during the melting at lower densities, where the slope of $T_m(P)$ is positive, and increases at higher densities where the function $T_m(P)$ decreases. Such behavior is consistent with the standard thermodynamic constraints that require lower density of the liquid phase in the region of positive slope of $T_m(P)$ and *vice versa*.

**V. CONCLUSIONS**

Turnover of the melting line detected within the framework of a simple AAP model is, in our opinion, the most important qualitative finding of this work. The fact that this effect is manifested under a single interaction model, contradicts the assumption of Bonev *ea*.[1] that for the melting line turnover are responsible certain changes in the intermolecular interaction.

Summarizing the above discussion, one may conclude that the reason for the turnover effect in hydrogen is a kind of frustration experienced by non-spherical molecules placed in sites of a crystal lattice. This frustration is related to deformation of chemical bonds progressing with increasing density. The short-range atom-atom repulsion hinders molecular rotation and leads eventually to the progressive shortening of chemical bonds and to the increase of vibronic frequency and intramolecular energy in solid phase when the pressure increases.

According to our simulations, the average bond length is shortened with increasing pressure both in solid and liquid phases but it always elongates noticeably on melting. Energy stored in frustrated chemical bonds is released during melting and this effect becomes more pronounced at higher densities and leads eventually to the progressive decrease of the stability of crystalline lattice, of melting temperature and vibronic frequency.

These conclusions based on our particular simulations results could have much more general scope. Of course, AAP approximation cannot pretend on precise description of interactions between molecules both at low and extremely high densities and temperatures. Nevertheless, comparing interaction energy predicted by AAP with the results of direct quantum-mechanical calculations for $H_2$-$H_2$ system and experiments on the scattering of molecular beams[9], one can see that this model gives a quite satisfactory overall description of the short-range repulsion of $H_2$ molecules.

At the same time, we should note that at large intermolecular distances AAP model does not recover the asymptotic behavior of the long-ranged orientational part of the intermolecular potential, in particular, of its quadrupole-quadrupole component. This may be the cause of an essential underestimation of the melting temperature at relatively low densities.

It should be also mentioned that at very short distances AAP approximation overestimates intermolecular repulsion. Pressure and ground state energy of solid hydrogen predicted by AAP model exceeds the results of more sophisticated calculations and measurements[6] at maximal density of 1.0 g cm$^{-3}$ up to two times.

These issues may be important at very high densities, where the pressure ionization effects predicted earlier[18] were recently observed at room temperature[19] and explained on the basis of the corresponding generalization of AAP model on ionic systems[20]. In any case, they do not affect the existence of the melting line turnover in classical molecular hydrogen described above.

**Table I.** Adopted MD simulation parameters.

| | |
|---|---|
| Number of atoms in the box | 512 or 1782 |
| Cut-off radius of non-valent potential, $Å$ | 6.5 |
| Time step, $fs$ | 0.125 |
| Equilibration period, $ps$ | 10 |
| Maximum period of monitoring, $ps$ | 300 |
| Period of recording of averaged values, $ps$ | 0.25 |
| Periodicity of the JMol-file appending, $fs$ | 2.5 - 250 |
| Set of densities studied, $g\ cm^{-3}$ | 0.3, 0.4, 0.5, 0.6, 0.7, 0.75, 0.775, 0.7875, 0.8, 0.8125, 0.825, 0.85, 0.875, 0.9, 0.925, 0.95 |

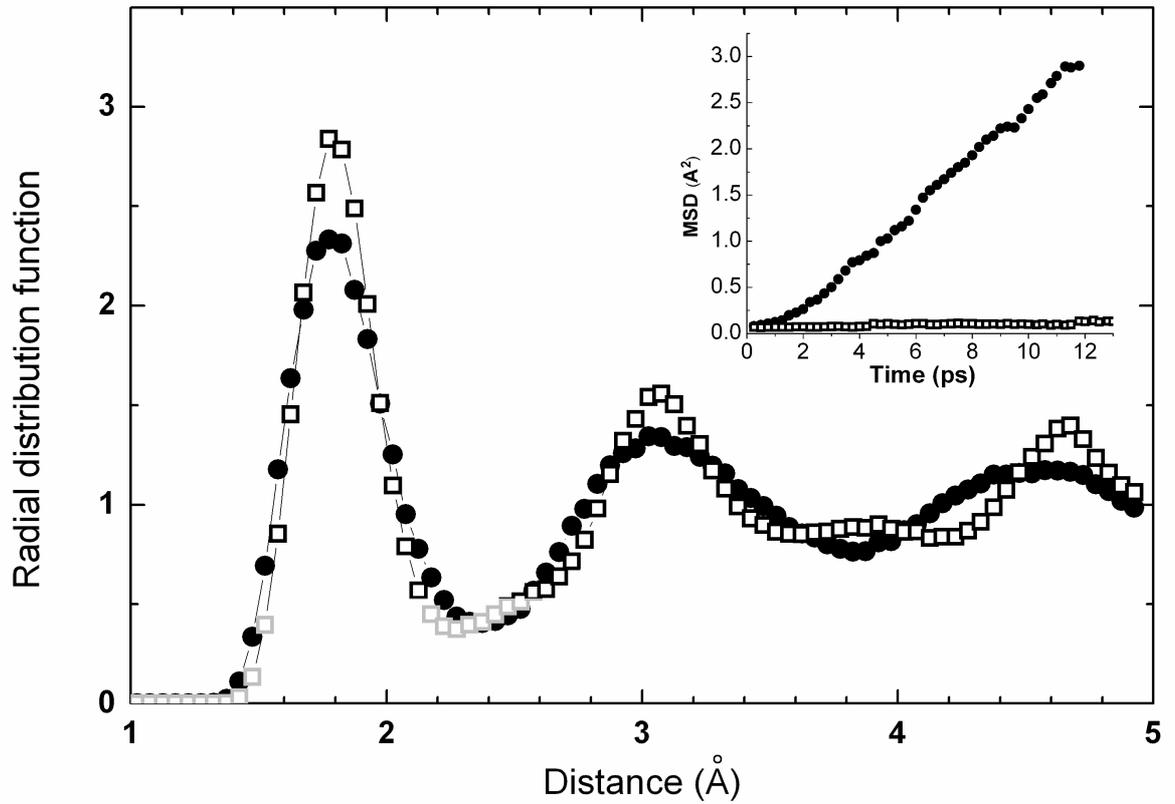

**Fig.1.** Changes in molecular center-to-center radial distribution function and mean squared displacements (inset) indicating melting of hydrogen at density 0.875 g cm$^{-3}$ ( P = 265 GPa, 512 atoms in box). Squares (red online) correspond to the solid phase (T=600 K), circles indicate liquid (T=625 K).

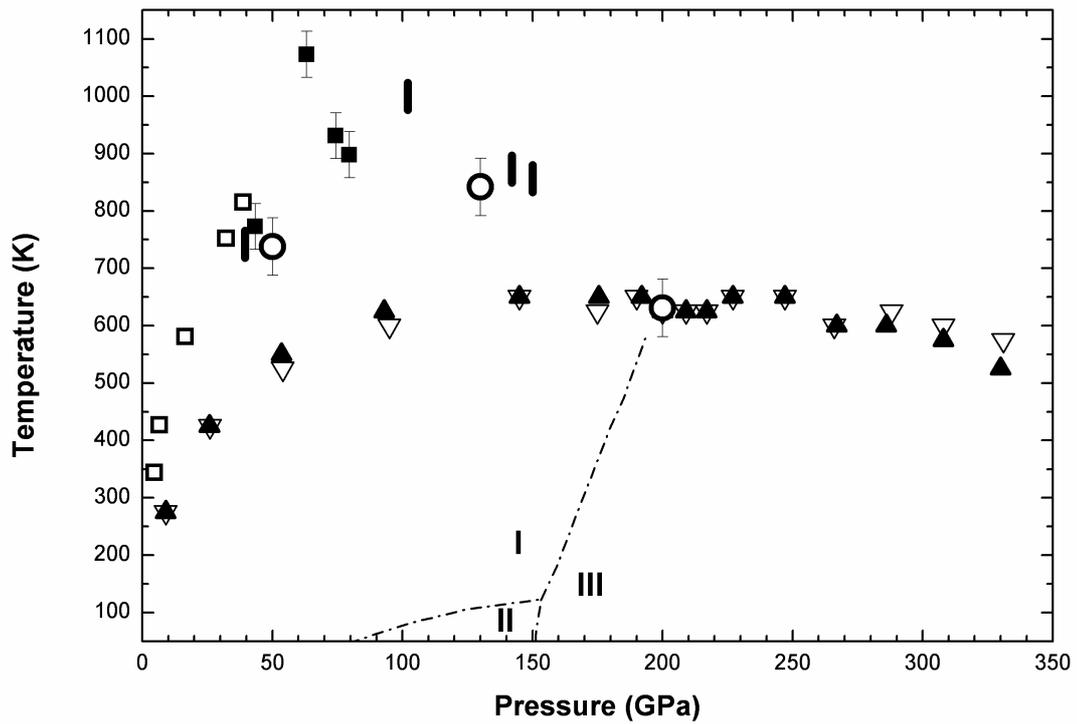

**Fig. 2.** Comparison of predicted melting temperatures with experimental and *ab initio* data. Results of MD simulations in a big box (1728 atoms) are marked by solid triangles, open triangles indicate the same in the smaller box (512 atoms). *Ab initio* results of Bonev ea.[1] shown by open circles (red online). Experimental data of Eremets and Troyan[3] shown by black vertical bars, measurements of Gregoryanz ea.[16] presented by open squares (blue online), data of Deemyad and Silvera[2] marked by black solid squares. Location of low-temperature polymorph I-II, II-II and I-III transitions in solid phase is indicated by dot-dashed lines according to Silvera and Deemyad[21].

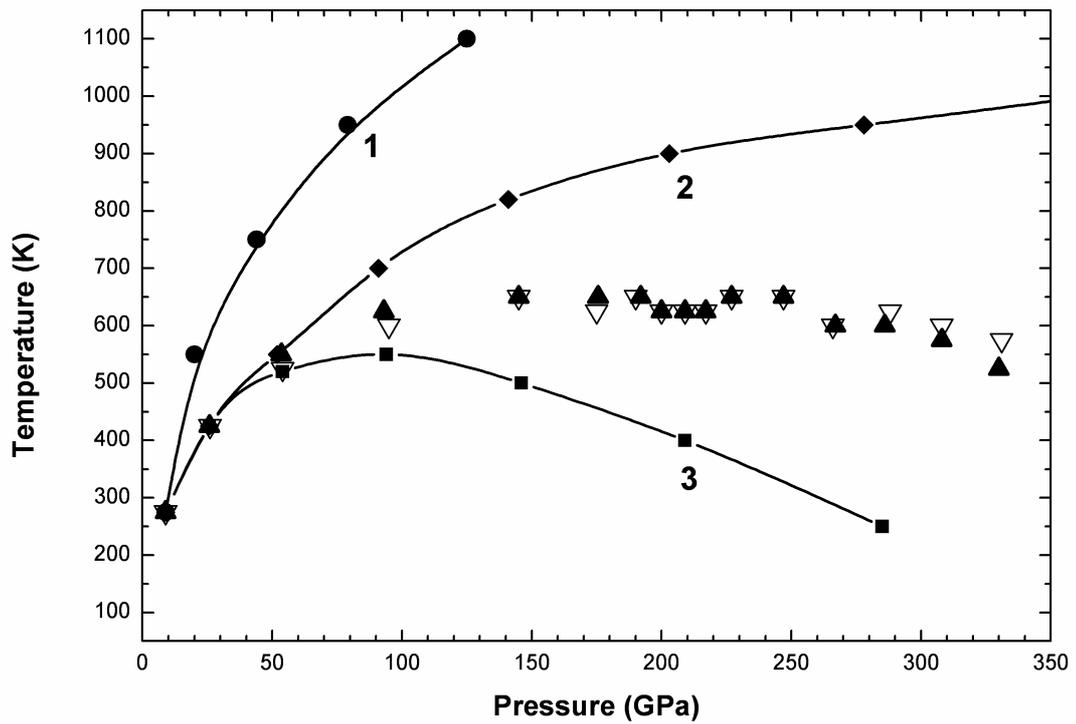

**Fig.3.** Effect of chemical bonding factor on the melting curve of hydrogen. Triangles represent the same MD results as in Fig.2. Three additional melting lines predicted using small MD box (512 atoms) and the following parameters of chemical bonding in Eq.(3) correspond to:

1 – equilibrium bond length $r_e$ shortened by half – (circles, red online);

2 – stiffness parameter $b$ reduced by half – (diamonds, brown online);

3 – doubled stiffness parameter $b$ (squares, green online);